# Standards for Energy Efficient Virtualization, Content Distribution and Big Data in Beyond 5G Networks

Jaafar M.H. Elmirghani, Hatem Alharbi, Azza Eltraify, Sanaa Hamid Mohamed, Barzan A. Yosuf and Taisir E.H. El-Gorashi
School of Electronic and Electrical Engineering, University of Leeds, UK

**Abstract**
Power consumption in communication networks and the supporting computing systems continues to increase due to the increase in traffic and processing requirements, and due to the relatively slower improvements in energy efficiency. Future networks are expected to continue to move computing algorithms and capabilities into the network including increased use of analytics, machine learning and intelligence applied to big data in the network, with content caching and virtualization. This article summarizes the key features of five new IEEE standards currently being developed to improve the energy efficiency of networks beyond 5G.

**Introduction**
The traffic in access and core networks continues to grow driven by new applications of the Internet of Things (IoT), machine to machine communication, high definition video, high data rate mobile traffic - such as in 5G and beyond and the increasing use of machine learning and artificial intelligence applied to big data carried by the network [1]. Traffic is currently growing at 30% - 40% per year on average [1] and if the current trends are sustained, this can lead to traffic doubling every two years, increasing by 30x in 10 years and 1000x in 20 years. The power consumption of the networks and the clouds and edge processing in the network can increase therefore by corresponding amounts. This led to the 1000x energy efficiency challenge in future networks. The solution to this challenge was originally pioneered by the GreenTouch consortium of 50 industrial and academic member organizations [2]. If the networks energy efficiency can be improved by 1000x, then in 20 years, networks can consume the same amount of energy as today while carrying 1000x more traffic.

Concurrently, a number of tools migrated from computing to networking including virtualization, caching and analytics. These tools can be used to improve the energy efficiency of 5G networks and beyond. For example, content can be cached near end users to reduce the journeys up and down the network when accessing content, thus resulting in power savings [3]. Virtualization can help improve resource utilization, thus resulting in power savings [4] - [6]. Furthermore, analytics can help learn about the network and optimize its operation in addition to learning about the user and hence optimizing the services [7], [8]. Figure 1 shows a network architecture which includes the core network, metro network segments and the access networks supported. It provides processing capabilities in the access layer in the form of access fog units, in the metro layer in the form of metro fog and in central clouds connected to core network nodes.

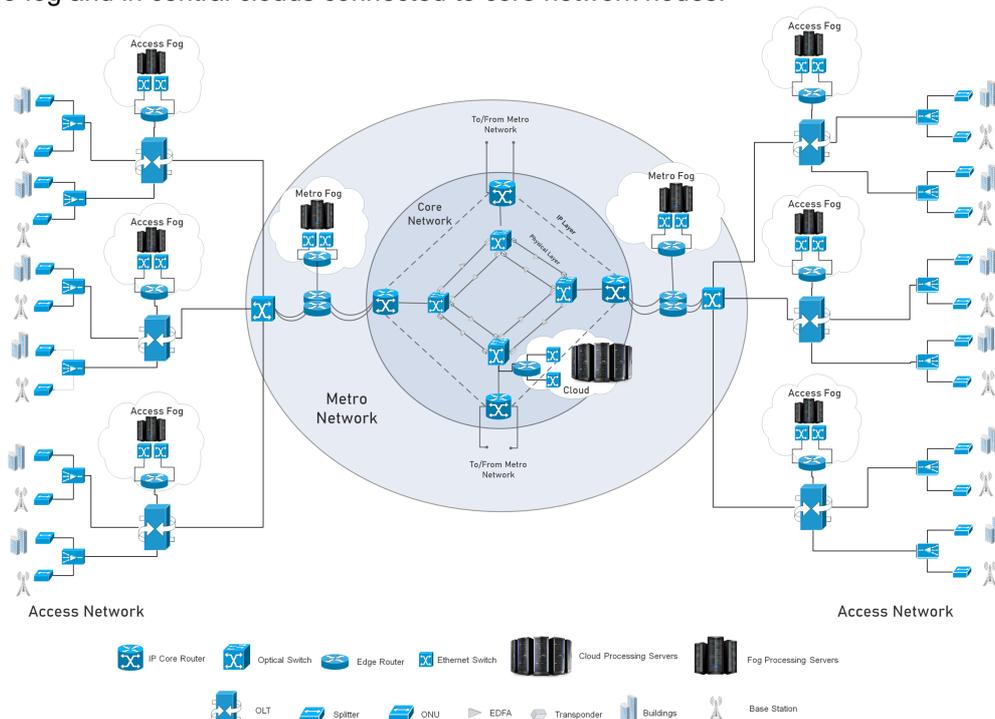

Fig. 1 Energy Efficient Network Virtualization and Content Distribution in Beyond 5G Access

The IEEE is developing a number of standards for energy efficient operation of networks in the presence of virtualization, content caching and big data analytics which can be understood by referring to the architecture in Fig. 1. The standards are IEEE P1925.1, IEEE P1926.1, IEEE P1927.1, IEEE P1928.1 and IEEE P1929.1 [9].

**The Standards**

User content generated at the edge of the network can aggregate and form big data streams. If such big data streams are transmitted to cloud data centres in the core network in Fig. 1 for processing, then significant power may be consumed. A key observation however is that users are generally interested in the knowledge embedded in the big data and not in the data stream itself. Consider an example where a heart rate monitor measures and transmits big data (due to its complex waveforms and multiple measurements) to a central data centre for processing. This data will hopefully repeatedly indicate that the person is fine for the next 30 years for example. Therefore, sending the full data may be redundant. Instead the big data stream can be processed at the edge of the network to extract incidents, ie knowledge, and transmit these incidents when the person needs help instead of transmitting the full data. The network therefore carries knowledge and not data. This form of big data edge processing can reduce latency and can save power by reducing the amount of data flowing in the network. This has to be done while paying attention to the key attributes of big data which include its Volume, its Velocity (meaning that it can be time sensitive and fast changing), its Variety (coming from multiple sources and sensors) and its Veracity [7], [8], [10]. The IEEE P1926.1 a "Standard for a Functional Architecture of Distributed Energy Efficient Big Data Processing" standardizes this form of big data edge processing.

The network in Fig. 1 can be virtualized to provide network slices to different users and different applications in an energy efficient manner. A user may request a network virtual slice that contains dedicated access to part of the capacity of core routers, core optical switches / multiplexers and demultiplexers, as well as dedicated access to part of the capacity of the fibre links interconnecting these nodes. In addition, the slice may contain certain computational and storage resources at the core network clouds and further computational and storage resources at the metro fog and access fog and the interconnecting networks [4] - [6]. Orchestrating these network resources in an energy efficient manner is the subject of IEEE P1927.1 "Standard for Services Provided by the Energy-efficient Orchestration and Management of Virtualized Distributed Data Centers Interconnected by a Virtualized Network" standard.

In several scenarios user computations may be carried out in virtual machines. It is usually preferable to place such virtual machines in the architecture in Fig. 1 as close to the end user as possible [2], [5]. For example, in the access fog processing units. As time progresses, users in a different part of the network may become interested in the services provided by this virtual machine while the original users may no longer be interested in the virtual machine due to time differences between the two user groups and work / entertainment switch over. In this case the virtual machine can be migrated to near the new set of users. In other cases, users interested in a given virtual machine may be at remote ends of the network. In such a case it may be necessary to replicate the virtual machines in a manner that minimizes power consumption. Finally, the hardware over which the virtual machine is implemented may have power consumption that is proportional to the load placed on the virtual machine. Here, it may be possible to slice the virtual machine and make several copies of the virtual machine, where all the copies collectively consume an amount of power equal to that consumed by one large virtual machine that runs all the tasks. This optimized, energy efficient placement of virtual machines is the subject of IEEE P1928.1, a "Standard for a Mechanism for Energy Efficient Virtual Machine Placement".

Video represents by some estimates over 80% the overall traffic volume in networks. A large part of the video traffic consumed is attributed to a small number of very popular videos. For instance, a video library that has few hundred million videos, may have 50 to 100 videos that account for 80 percent of the requests, the well known heavy tail Zipf distribution. If these few videos are cached locally, in the access fog or metro fog in Fig. 1 instead of being stored in the central cloud, then the network power consumption associated with retrieving these videos from the central library is decreased. The addition of network video caching hardware increases the overall hardware power consumption. This is tensioned against the power saved as a result of caching. As such an optimum video cache size exists that results in power minimization in the network [2], [3]. The IEEE P1929.1 standard "An Architectural Framework for Energy Efficient Content Distribution" addresses this problem.

Finally, in many network segments, the traffic may not fill a transponder or alternatively may just exceed the capacity of a single transponder, which calls for the use of two transponders. In both cases the unused transponder capacity leads to energy wastage. An alternative is to use mixed line rates where transponders operating at different data rates are used. These may include transponders operating at 10 Gb/s, 40 Gb/s, 100

Gb/s, 400 Gb/s and beyond. There is thus a correct combination of transponders that results in power consumption minimization. A similar result can be achieved if the transponders used adaptive line rates based on optical orthogonal frequency division multiplexing (OFDM) [2], [11]. The IEEE P1925.1 "Standard for Energy Efficient Dynamic Line Rate Transmission System" standard focuses on the energy efficiency of these systems.

**Conclusions and future work**
A number of standards are being developed to improve the energy efficiency of future big data networks, content distribution networks, and virtualized networks. The standards take into account the network architecture, the hardware and the algorithms needed to orchestrate, manage and provision the networks. Future work will address the expected increased use of machine learning, and intelligence in networks to improve energy efficiency, and to provide a better fit between the users, services and network.

**Acknowledgements**
We would like to acknowledge funding from the Engineering and Physical Sciences Research Council (EPSRC), INTERNET (EP/H040536/1), STAR (EP/K016873/1) and TOWS (EP/S016570/1) projects.

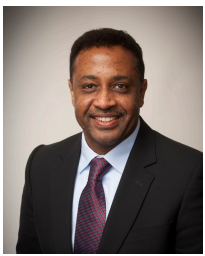

**Prof. Jaafar Elmirghani** is FIET, FIoP, and Director of the Institute of Integrated Information Systems, Leeds. He has provided outstanding leadership in a number of large research projects, and was PI of the £6m EPSRC Intelligent Energy Aware Networks (INTERNET) Programme Grant, 2010-2016. He is Co-Chair of the IEEE Sustainable ICT initiative, a pan IEEE Societies initiative responsible for Green ICT activities across IEEE. He was awarded the IEEE Comsoc 2005 Hal Sobol award, 2 IEEE Comsoc outstanding service awards (2009, 2015), the 2015 GreenTouch 1000x award, IET Optoelectronics 2016 Premium Award and shared the 2016 Edison Award in the collective disruption category with a team of 6 from GreenTouch for joint work on the GreenMeter. His work led to 5 IEEE standards with a focus on network virtualisation and energy efficiency, where he currently heads the work group responsible for IEEE P1925.1, IEEE P1926.1, IEEE P1927.1, IEEE P1928.1 and IEEE P1929.1, this resulting in significant impact through industrial and academic uptake. He is PI of the EPSRC £6.6m Terabit Bidirectional Multi-user Optical Wireless System (TOWS) for 6G LiFi, 2019-2024. He was an IEEE

Comsoc Distinguished Lecturer 2013-2016. He has published over 500 technical papers, and has research interests in communication systems and networks energy efficiency and in optical wireless systems and networks.